\documentclass{sig-alternate-05-2015}

\permission{Copyright is held by the International World Wide Web Conference Committee (IW3C2). IW3C2 reserves the right to provide a hyperlink to the author's site if the Material is used in electronic media.}
\conferenceinfo{WWW 2016 Companion,}{April 11--15, 2016, Montreal, Canada.} 
\copyrightetc{ACM \the\acmcopyr}
\crdata{123-4-5678-9012-3/45/67. \\
http://dx.doi.org/12.3456/7890123.4567890}

\usepackage{graphicx}
\usepackage{caption}
\usepackage{subcaption}
\usepackage{rotating}

\usepackage[utf8]{inputenc}
\usepackage{enumitem}
\usepackage{textcomp}

\PassOptionsToPackage{hyphens}{url}
\usepackage[pdftex,urlcolor=black,colorlinks=true,linkcolor=black,citecolor=black]{hyperref}

\usepackage[hyphens]{url}

\usepackage[usenames,dvipsnames,svgnames,table]{xcolor}
\usepackage{fancyvrb}
\usepackage{relsize}
\usepackage{listings}
\usepackage{verbatim}
\newcommand{\defaultlistingsize}{\fontsize{8pt}{9.5pt}}
\newcommand{\inlinelistingsize}{\fontsize{8pt}{11pt}}

\newcommand{\listingsize}{\defaultlistingsize}
\RecustomVerbatimCommand{\Verb}{Verb}{fontsize=\inlinelistingsize}
\RecustomVerbatimEnvironment{Verbatim}{Verbatim}{fontsize=\defaultlistingsize}
\let\oldurl\url
\renewcommand{\url}[1]{\inlinelistingsize\oldurl{#1}}

\lstdefinelanguage{JavaScript}{
  keywords={console, log, addEventListener, onmessage, alert, push, typeof, new, true, false, catch, function, return, null, catch, switch, var, if, in, while, do, else, case, break},
  keywordstyle=\bfseries,
  ndkeywords={class, export, boolean, throw, implements, import, this},
  ndkeywordstyle=\color{darkgray}\bfseries,
  identifierstyle=\color{Maroon},
  sensitive=false,
  comment=[l]{//},
  morecomment=[s]{/*}{*/},
  commentstyle=\color{ForestGreen},
  stringstyle=\color{Blue},
  morestring=[b]',
  morestring=[b]"
}

\usepackage{color}
\definecolor{grey}{RGB}{130,130,130}

\usepackage{color}

\begin{document}

\title{Wikipedia Tools for Google Spreadsheets}

\numberofauthors{1}

\author{
\alignauthor
Thomas Steiner\\
       \affaddr{Google Germany GmbH}\\
       \affaddr{ABC Str. 19, 20354 Hamburg, Germany}\\
       \email{\fontsize{12pt}{14.4pt}\sffamily\selectfont tomac@google.com}
}

\maketitle
\begin{abstract}
In this paper, we introduce the \emph{Wikipedia Tools for Google Spreadsheets}.
Google Spreadsheets is part of a~free, Web-based software office suite
offered by Google within its Google Docs service.
It allows users to create and edit spreadsheets online,
while collaborating with other users in realtime.
Wikipedia is a~free-access, free-content Internet encyclopedia,
whose content and data is available, among other means, through an API.
With the \emph{Wikipedia Tools for Google Spreadsheets}, we have created a~toolkit
that facilitates working with Wikipedia data from within a~spreadsheet context.
We make these tools available as open-source on GitHub,%
\footnote{\emph{Wikipedia Tools for Google Spreadsheets}:
\url{https://github.com/tomayac/wikipedia-tools-for-google-spreadsheets}}
released under the permissive Apache~2.0 license.
\end{abstract}



\section{Introduction}

In the world of Computer Science, \emph{spreadsheet} applications
serve for the organization, analysis, and storage of data in tabular form.
Spreadsheets are the computerized simulation of paper accounting worksheets,
and operate on data represented as \emph{cells of an array},
organized in rows and columns.
Cells can contain numeric or textual data, or the results of \emph{formulas}
that automatically calculate and display a~value based on the contents of other cells.
With the \emph{Wikipedia Tools for Google Spreadsheets},
we introduce a~toolkit of such formulas, tailored to the universe of Wikipedia,
that enables a~wide range of potential use cases
starting from marketing, to search engine optimization, to business analysis.
Especially through the \emph{chaining} of formulas, the true power and ease of spreadsheet
applications can be unleashed.

\subsection{Wikipedia and Wikidata}

Wikipedia's content and data is available through the Wikipedia API
(\url{https://{language}.wikipedia.org/w/api.php}),
where \texttt{\{language\}} represents one of the currently 291~supported Wikipedia languages,%
\footnote{List of Wikipedias:
\url{https://meta.wikimedia.org/wiki/List_of_Wikipedias}}
for example, \texttt{en} for English, \texttt{de} for German, or \texttt{zu} for Zulu. 
Wikidata is a~collaboratively edited knowledge base and intended to provide
a~common source of structured data which can be used by projects such as Wikipedia.
Its content and data is available through the Wikidata API
(\url{https://www.wikidata.org/w/api.php}).
Both the Wikipedia and the Wikidata APIs' data is available as XML or JSON, among other formats.
Wikipedia pageviews data, \emph{i.e.}, the number of times within a~given period of time
that a~given Wikipedia article has been viewed can be obtained using the Pageviews API
(\url{https://wikimedia.org/api/rest_v1/?doc}).
The data is available in JSON format.

\subsection{Google Spreadsheets and Apps Scripts}

Google Spreadsheets can be extended with custom functions (or formulas)
using Google Apps Scripts\footnote{Google Apps Script:
\url{https://developers.google.com/apps-script/}}
that are written in standard JavaScript.%
\footnote{Custom functions in Google Sheets:
\url{https://developers.google.com/apps-script/guides/sheets/functions}}
To illustrate this, a~trivial function is defined in \autoref{code:custom-function-definition}
that can then be used from within a~spreadsheet as outlined in
\autoref{code:custom-function-usage}.
Custom functions can access external resources on the Web by fetching URLs
with the \texttt{UrlFetchApp}, one of the scripting services available in
Google Apps Script.
Fetched data can either be in XML or JSON format and parsed with convenience functions.

\vspace{-1.5em}
\begin{figure}[h!]
\begin{lstlisting}[caption={Custom Google Sheets function called \texttt{DOUBLE}.},
  label=code:custom-function-definition, language=JavaScript]
function DOUBLE(input) {
  return input * 2;
}
\end{lstlisting}
\vspace{-2.5em}
\begin{lstlisting}[caption={Usage of the custom \texttt{DOUBLE} function from
  \autoref{code:custom-function-definition} in a~cell with the value of cell \texttt{A1}}
  as a~parameter.,
  label=code:custom-function-usage, language=JavaScript]
=DOUBLE(A1)
\end{lstlisting}
\end{figure}

\vspace{-2.5em}
\section{List of Developed Functions}

\begin{sidewaysfigure}[p!]
  \centering
  \includegraphics[width=\textwidth]{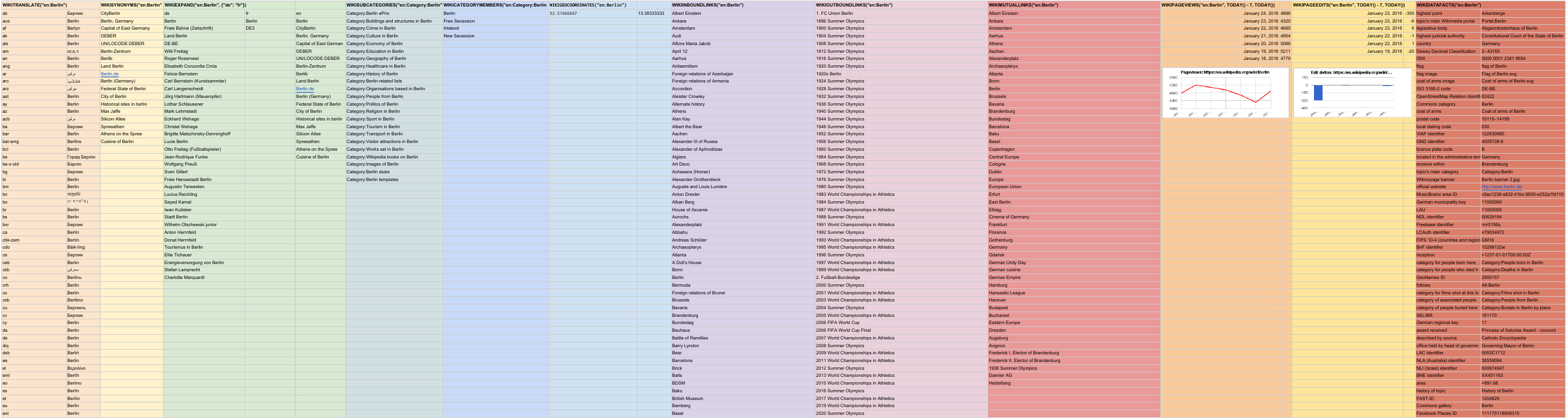}
  \caption{Example output for each function in the \emph{Wikipedia Tools for Google Spreadsheets} (cropped). Live spreadsheet: \url{https://goo.gl/yvbmex}.}
  \label{fig:wikipedia-tools}
\end{sidewaysfigure}

In our \emph{Wikipedia Tools for Google Spreadsheets}, we provide
eleven functions that---in traditional spreadsheets style---%
follow an all-uppercase naming convention and start with a~\texttt{WIKI} prefix.
These functions are wrappers around the particular Wikipedia or Wikidata API calls,
or the Pageviews API respectively.
\autoref{fig:wikipedia-tools} shows exemplary output for the English Wikipedia article
\url{https://en.wikipedia.org/wiki/Berlin} and the English Wikipedia category
\url{https://en.wikipedia.org/wiki/Category:Berlin}.
The functions are listed below.

\begin{description}[style=unboxed,leftmargin=1em]
  \itemsep0em
  \item[\texttt{WIKITRANSLATE}] Returns Wikipedia translations (language links) for a~Wikipedia article.
  \item[\texttt{WIKISYNONYMS}] Returns Wikipedia synonyms (redirects) for a~Wikipedia article.
  \item[\texttt{WIKIEXPAND}] Returns Wikipedia translations (language links) and synonyms (redirects) for a~Wikipedia article.
  \item[\texttt{WIKICATEGORYMEMBERS}] Returns Wikipedia category members for a~Wikipedia category.
  \item[\texttt{WIKISUBCATEGORIES}] Returns Wikipedia subcategories for\linebreak a~Wikipedia category.
  \item[\texttt{WIKIINBOUNDLINKS}] Returns Wikipedia inbound links for\linebreak a~Wikipedia article.
  \item[\texttt{WIKIOUTBOUNDLINKS}] Returns Wikipedia outbound links for a~Wikipedia article.
  \item[\texttt{WIKIMUTUALLINKS}] Returns Wikipedia mutual links, i.e, the intersection of inbound and outbound links for a~Wikipedia article.
  \item[\texttt{WIKIGEOCOORDINATES}] Returns Wikipedia geocoordinates for a~Wikipedia article.
  \item[\texttt{WIKIDATAFACTS}] Returns Wikidata facts for a~Wikipedia\linebreak article.
  \item[\texttt{WIKIPAGEVIEWS}] Returns Wikipedia pageviews statistics for a~Wikipedia article.
  \item[\texttt{WIKIPAGEEDITS}] Returns Wikipedia pageedits statistics for a~Wikipedia article.
\end{description}

Most functions directly wrap native API calls, with three exceptions:
\emph{(i)} the functionality of the \texttt{WIKISYNONYMS} and the \texttt{WIKITRANSLATE} 
functions is combined in the \texttt{WIKIEXPAND} function,
both the \texttt{WIKITRANSLATE} and the \texttt{WIKIEXPAND} function
accept an optional target languages parameter that allows for limiting
the output to just a~subset of all available Wikipedia languages;
\emph{(ii)} the function \texttt{WIKIMUTUALLINKS} is the intersection
of the two functions \texttt{WIKIINBOUNDLINKS} and \texttt{WIKIOUTBOUNDLINKS};
and \emph{(iii)} the function \texttt{WIKIDATAFACTS} provides
a~list of claims~\cite{vrandecic2014wikidata} (or facts), enriched
with entity and property labels for improved readability,
limited to single-value objects, and simplified using an adapted version of Maxime Lathuilière's
\texttt{simplifyClaims} function%
\footnote{Wikidata SDK \texttt{simplifyClaims} function:
\url{https://github.com/maxlath/wikidata-sdk\#simplify-claims-results}}
from his Wikidata SDK~\cite{lathuiliere2016wikidatasdk}.
This allows us to return two columns%
---in RDF~\cite{cyganiak2014rdf} terms ``predicate'' and ``object'' pairs---%
with one unique object, for example, the predicate \texttt{ISO 3166-2 code}
with the object \texttt{DE-BE},
and deliberately discarding multi-value claims, for example,
predicate \texttt{head of government} with objects \texttt{Michael Müller} and
\texttt{Klaus Wowereit}, among many others.
While in the concrete example the ordering is clear (temporal),
this is not true in the general case, for example, with predicate \texttt{instance of}.
As a~result, in \texttt{WIKIDATAFACTS}, we prefer
indisputability of claims over their completeness.
\autoref{code:wikisynonyms} exemplarily shows the complete implementation
of the \texttt{WIKISYNONYMS} function.

\begin{lstlisting}[caption={Implementation of \texttt{WIKISYNONYMS}.},
  label=code:wikisynonyms, language=JavaScript]
/**
 * Returns Wikipedia synonyms
 * @param {string} article The Wikipedia article
 * @return {Array<string>} The list of synonyms
 */
function WIKISYNONYMS(article) {
  'use strict';
  if (!article) {
    return '';
  }
  var results = [];
  try {
    var language = article.split(/:(.+)?/)[0];
    var title = article.split(/:(.+)?/)[1];
    if (!title) {
      return '';
    }
    title = title.replace(/\s/g, '_');
    var url = 'https://' + language +
        '.wikipedia.org/w/api.php' +
        '?action=query' +
        '&blnamespace=0' +
        '&list=backlinks' +
        '&blfilterredir=redirects' +
        '&bllimit=max' +
        '&format=xml' +
        '&bltitle=' +
        encodeURIComponent(title);
    var xml = UrlFetchApp.fetch(url)
        .getContentText();
    var document = XmlService.parse(xml);
    var entries = document.getRootElement()
        .getChild('query').getChild('backlinks')
        .getChildren('bl');
    for (var i = 0; i < entries.length; i++) {
      var text = entries[i].getAttribute('title')
          .getValue();
      results[i] = text;
    }
  } catch (e) {
    // no-op
  }
  return results.length > 0 ? results : '';
}
\end{lstlisting}

\section{Usage Scenarios}

We have tested the \emph{Wikipedia Tools for Google Spreadsheets}
with different usage scenarios in mind.
These include, but are not limited to, the ones listed in the following.

\subsection{Usage Scenario I: Ordered Category Panel}

\begin{figure}[p!]
  \centering
  \includegraphics[width=0.3\textwidth]{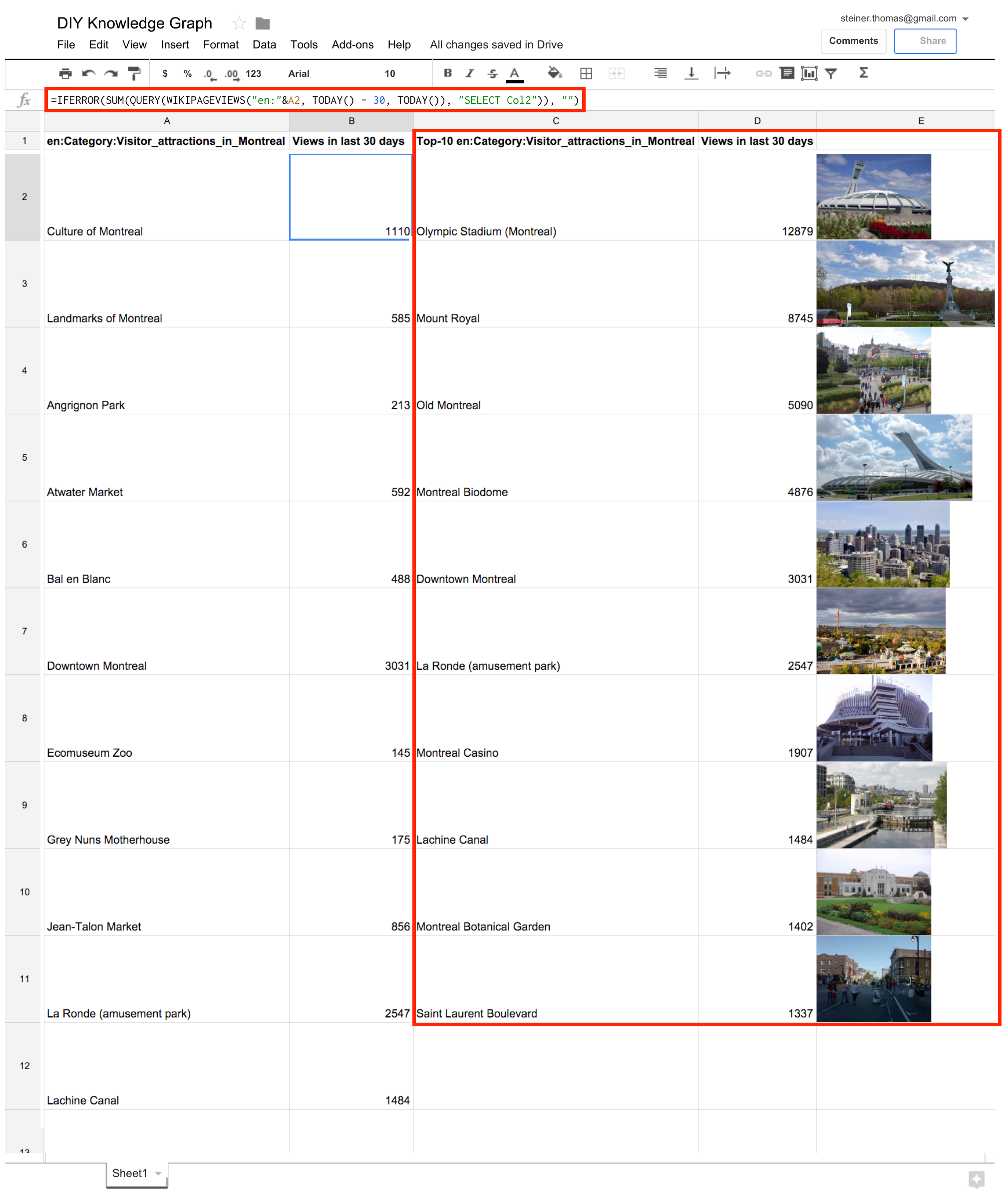}
  \caption{Usage scenario I: \emph{Wikipedia Tools for Google Spreadsheets} used to create an ordered category panel based on Wikipedia category memberships and accumulated Wikipedia pageviews for popularity ranking (here: the top-10 visitor attractions in Montreal). Live spreadsheet: \url{https://goo.gl/Njvt1T}.}
  \label{fig:knowledge-graph}
\end{figure}

Wikipedia holds an enormous amount of categories, for example,
\emph{visitor attractions in Montreal}.%
\footnote{Visitor attractions in Montreal:
\url{https://en.wikipedia.org/wiki/Category:Visitor_attractions_in_Montreal}}
Category members obtained through a~call of \texttt{WIKICATEGORYMEMBERS}
are listed in alphabetical order,
however, if we additionally request pageviews data for each category member through a~series of \texttt{WIKIPAGEVIEWS} calls
and then sort by pageviews in descending order,
we get a~representative list of top-$10$ visitor attractions---%
enriched with photos retrieved through calls of \texttt{WIKIDATAFACTS} filtered on ``image''---%
as shown in \autoref{fig:knowledge-graph}.
A~similar feature (based on non-disclosed metrics) in form of an image carousel
can be seen in Google's Knowledge Graph~\cite{singhal2012} Web search results pages
when searching for ``visitor attractions in montreal'' (demo \url{https://goo.gl/Ugt0je}).

\subsection{Usage Scenario II: Search Ads}

\begin{figure}[p!]
  \centering
  \includegraphics[width=0.4\textwidth]{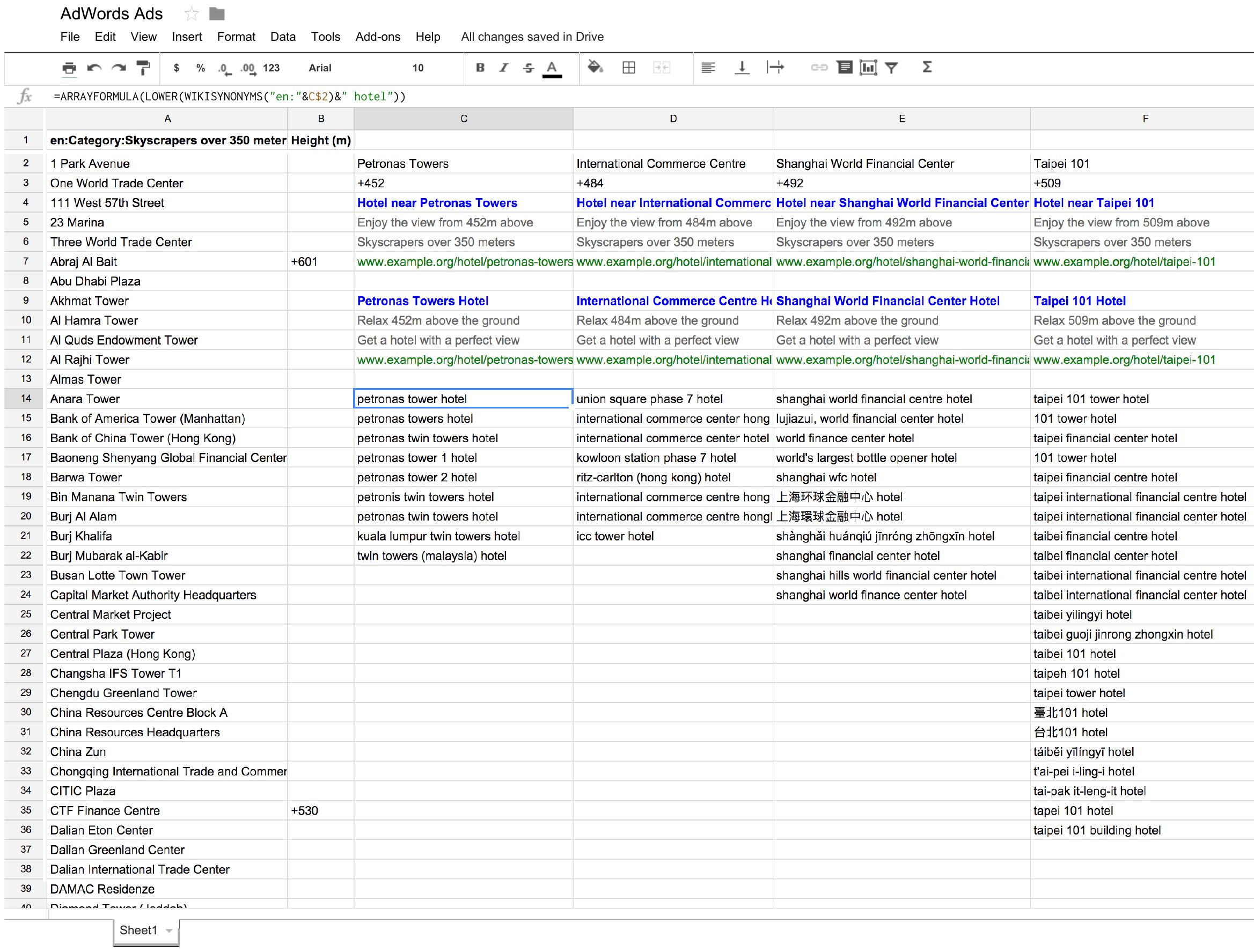}
  \caption{Usage scenario II: \emph{Wikipedia Tools for Google Spreadsheets} used to create textual search ads based on Wikidata facts (here: skyscraper heights) and Wikipedia synonyms as keywords combined with the term ``hotel''. Live spreadsheet: \url{https://goo.gl/np1Is8}.}
  \label{fig:adwords-ads}
\end{figure}

Search advertisers can greatly profit from the information
that is contained in Wikipedia and Wikidata.
For example, if we imagine a~hotel booking site,
it may be desirable to advertise based on points of interest (POIs)
and create advertisements automatically featuring known facts of such POIs.
\autoref{fig:adwords-ads} shows an example where skyscrapers listed in the category
\emph{skyscrapers over 350 meter}%
\footnote{Skyscrapers over 350 meter:
\url{https://en.wikipedia.org/wiki/Category:Skyscrapers_over_350_meters}}
 are first obtained via \texttt{WIKICATEGORYMEMBERS}
and then checked for their ``height'' fact via \texttt{WIKIDATAFACTS},
which is then used in two templates to create ads.
Search keywords are generated by calling \texttt{WIKISYNONYMS} and combined with terms like ``hotel''.

\subsection{Usage Scenario III: Marketing Campaigns}

\begin{figure}[p!]
  \centering
  \includegraphics[width=0.4\textwidth]{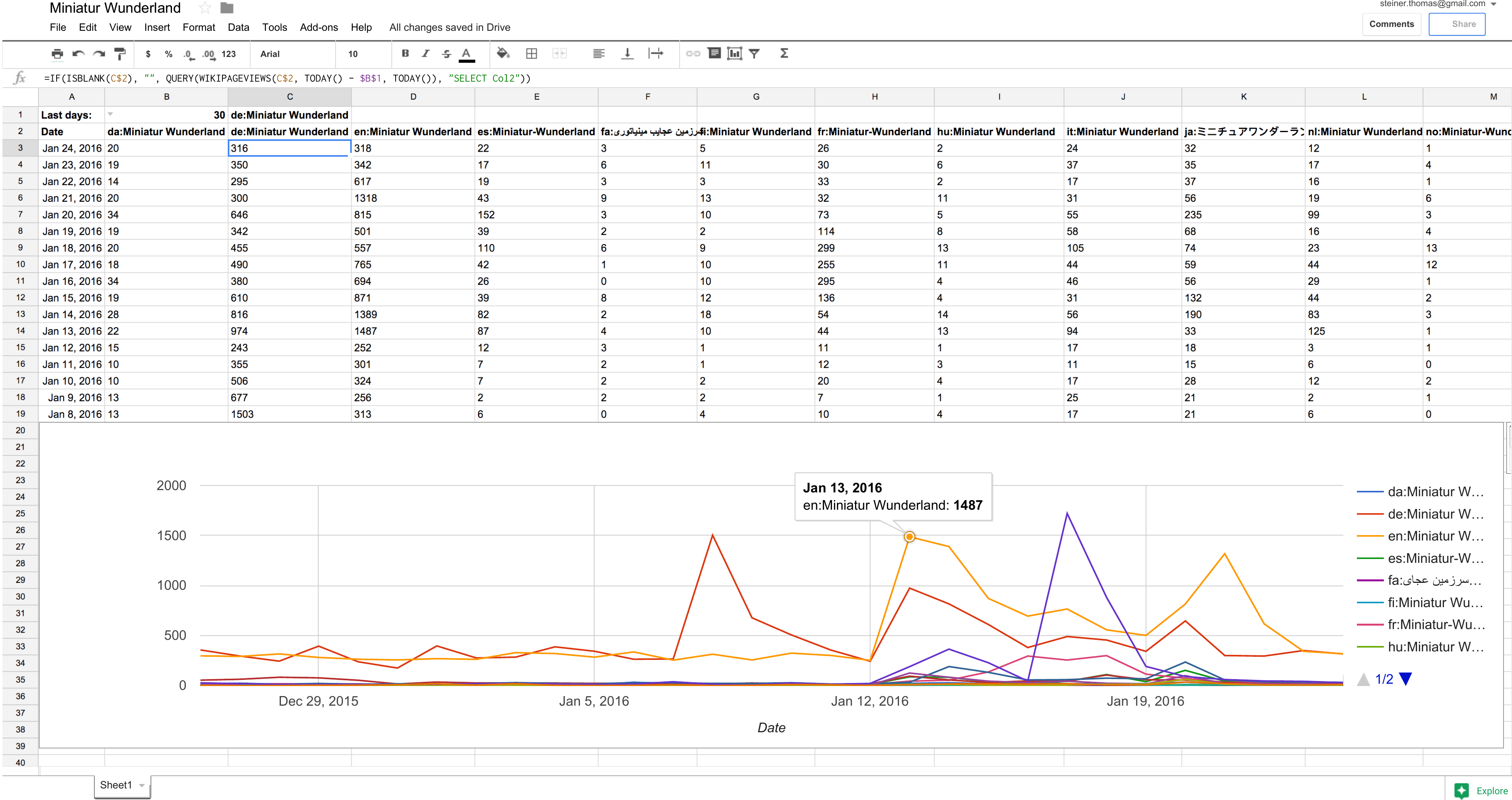}
  \caption{Usage scenario III: \emph{Wikipedia Tools for Google Spreadsheets} used to evaluate the impact of a~marketing campaign (here: model railway \emph{Miniatur Wunderland} being featured on Google Street View since January~13, 2016). Live spreadsheet: \url{https://goo.gl/q1yhuV}.}
  \label{fig:miniatur-wunderland}
\end{figure}

On January~13, 2016, Google Maps added Street View imagery for the model railway \emph{Miniatur Wunderland}.%
\footnote{Miniatur Wunderland on Google Street View:
\url{https://www.google.com/maps/about/behind-the-scenes/streetview/treks/miniatur-wunderland/}}
Taking global Wikipedia pageviews as a~popularity indicator,
we can examine if the marketing campaign has had any impact on the attraction,
assuming that more pageviews translate to increased visitor interest.
Therefore, we first obtain the \emph{Miniatur Wunderland} article in all available languages
via \texttt{WIKITRANSLATE} and then retrieve pageviews via \texttt{WIKIPAGEVIEWS}.
\autoref{fig:miniatur-wunderland} shows indeed an international uptake of pageviews
starting January~13 after an earlier linear curve progression
(except for the German article, which had a~peak on January~8, a~long weekend after a~public holiday).

\section{Related Work}

In his book \emph{Google Apps Script for Beginners}~\cite{gabet2014google}, Gabet
gives an introduction to extending Google Spreadsheets with custom functions.
A~similar introduction is given in Ferreira's
\emph{Google Apps Script: Web Application Development Essentials}~\cite{ferreira2014google}.
In~\cite{han2008rdf123}, Han \emph{et~al.}\ describe their approach \emph{RDF123}
to translate spreadsheets data to RDF, the inverse of what we do in \texttt{WIKIDATAFACTS}.
Olsen and Moser show in~\cite{olsen2013teaching} how Web APIs can be taught with spreadsheets.
The process of calling Web APIs via spreadsheets is further described in~\cite{patel2014spreadsheet}.
Further, in~\cite{abramson2001supercomputing}, Abramson \emph{et~al.}\ describe
how they enabled spreadsheets to have ``super-computing'' powers through
parallelized custom functions.
An open-source toolkit for mining Wikipedia---not bound to spreadsheets,
but designed for general use with the Java programming language---%
is described by Milne \emph{et~al.}\ in~\cite{milne2013toolkit}.

\section{Conclusions and Future Work}

In this paper, we have introduced the \emph{Wikipedia Tools for Google Spreadsheets}.
First, we have introduced the data sour\-ces Wikipedia and Wikidata and their different APIs.
Second, we have shown how Google Spreadsheets can be extended through custom functions
that can then be used from within a~cell context as if they were native functions.
In the following, we have listed the implemented functions,
and explained where they extend the functionality of the underlying wrapped API functions.
We have then focused on three different usage scenarios that
illustrate how to work with the \emph{Wikipedia Tools for Google Spreadsheets}
and finally have provided an overlook on related work in the area.

Future work will focus on adding more functions as need be
and potentially making the functions more parameterizable.
In the current iteration, we have favored simplicity and ease of use over customizability,
essentially making the most common use case the only option.
Possibly, in upcoming releases, we will add an advanced mode
that allows experienced users to fine-tune the functions' results,
for example, to implicitly include bot traffic in \texttt{WIKIPAGEVIEWS}
that we have currently excluded on purpose.

Concluding, we were positively surprised by the increased productivity
and short turnaround time enabled by the \emph{Wikipedia Tools for Google Spreadsheets}
for the rapid prototyping of ideas, especially in combination with the fill-down and fill-right
features in spreadsheets and the charting capabilities.
We look forward to making the tools even more powerful
and hope to attract collaborators for the open source project
available on GitHub at \url{https://github.com/tomayac/wikipedia-tools-for-google-spreadsheets}.
As a~positive side effect, the tools can even help improve Wikipedia and Wikidata
when authors add missing data, for example, we added an image to one of the visitor attractions
of Montreal, as this fact was initially missing in Wikidata (and thus in \autoref{fig:knowledge-graph}).

\normalsize
\bibliographystyle{abbrv}
\bibliography{references}

\end{document}